# Development of a dual-spectroscopic system to rapidly measure diisopropyl methyl phosphonate (DIMP) decomposition and temperature in a reactive powder environment FREE

Preetom Borah ; Milad Alemohammad; Mark Foster ; Timothy P. Weihs

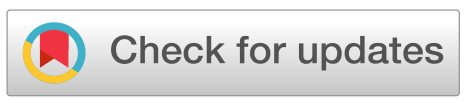

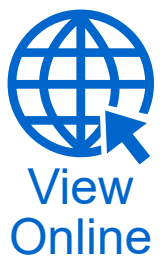

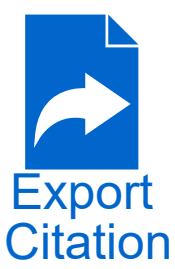

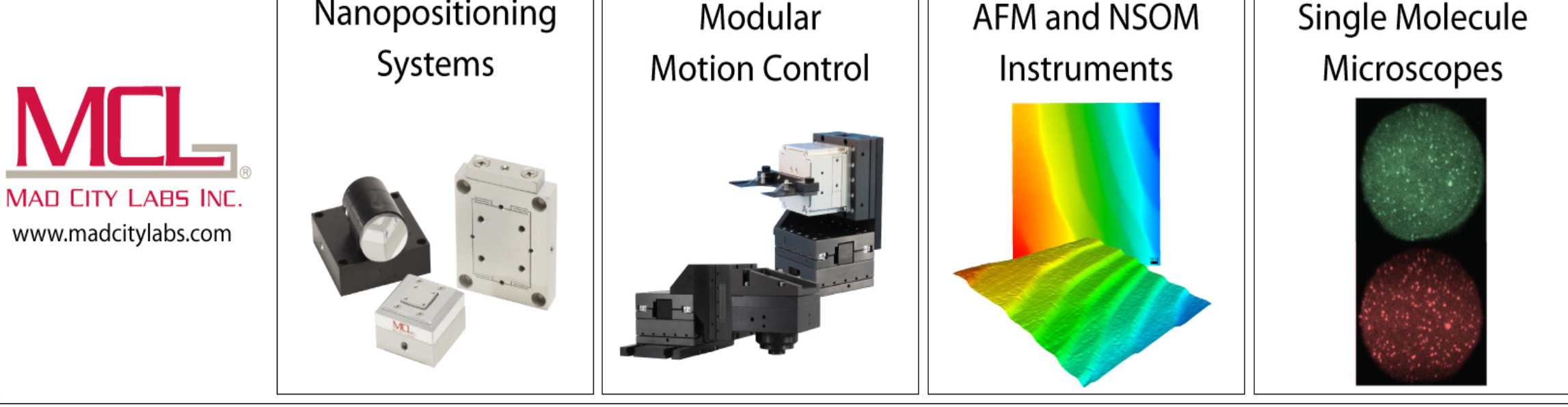

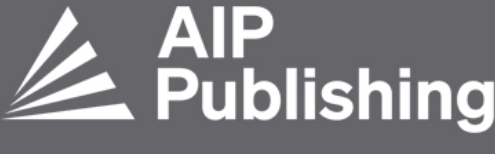

# Development of a dual-spectroscopic system to rapidly measure diisopropyl methyl phosphonate (DIMP) decomposition and temperature in a reactive powder environment

Preetom Borah,[1,2,a)] Milad Alemohammad,[2,3] Mark Foster,[2,3] 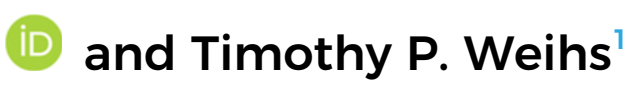 and Timothy P. Weihs[1,2,b)] 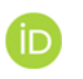

## AFFILIATIONS

[1] Department of Materials Science and Engineering, Johns Hopkins University, Baltimore, Maryland 21218, USA
[2] Hopkins Extreme Materials Institute, Johns Hopkins University, Baltimore, Maryland 21218, USA
[3] Department of Electrical and Computer Engineering, Johns Hopkins University, Baltimore, Maryland 21218, USA

[a)] **Author to whom correspondence should be addressed:** pborah1@jhu.edu
[b)] **Electronic mail:** weihs@jhu.edu

## ABSTRACT

The development of systems to measure and optimize emerging energetic material performance is critical for Chemical Warfare Agent (CWA) defeat. In order to assess composite metal powder efficacy on CWA simulant defeat, this study documents a combination of two spectroscopic systems designed to monitor the decomposition of a CWA simulant and temperature rises due to combusting metal powders simultaneously. The first system is a custom benchtop Polygonal Rotating Mirror Infrared Spectrometer (PRiMIRS) incorporating a fully customizable sample cell to observe the decomposition of Diisopropyl Methyl Phosphonate (DIMP) as it interacts with combusting composite metal particles. The second is a tunable diode laser absorption spectroscopy (TDLAS) used to monitor increases in background gas temperatures as the composite metal powders combust. The PRiMIRS system demonstrates a very high signal to noise ratio (SNR) at slow timescales (Hz), reasonable SNR when operating at faster timescales (100 Hz), and capabilities of resolving spectral features with a FWHM resolution of 15 $cm^{-1}$. TDLAS was able to monitor temperature rises between room temperature and 230 ± 5 °C while operating at 100 Hz. For testing, liquid DIMP was inserted in a preheated stainless steel (SS) cell to generate DIMP vapor and (Al–8Mg):Zr metal powders were ignited in a SS mount with a resistively heated nichrome wire at one end of the cell. The ignited particles propagated across the cell containing DIMP vapor. The path averaged gas temperature in the preheated SS cell rises rapidly (100 ms) and decays slowly (<5 s) but remains below 230 °C during particle combustion, a temperature at which the thermal decomposition of DIMP is not observed over similarly short timescales (seconds). However, when combusting particles were introduced to the DIMP vapor (heterogeneous environment), spectral signatures indicative of decomposition product formation, such as isopropyl-methyl phosphonate (IMP) and isopropyl alcohol, were observed within seconds.

## I. INTRODUCTION

Chemical warfare agents (CWAs) are of growing concern, given their recent use among civilian populations in Syria in 2018[1,2] and in Japan between 1994 and 1995.[3] Organophosphate nerve agents inhibit acetylcholinesterase activity within the nervous systems, resulting in brutal symptoms on victims; both uses resulted in thousands of injuries and deaths.[3] To help prevent future attacks, materials and methods are being developed to promptly neutralize CWAs and biological warfare agents (BWAs) prior to their use. One such method involves the use of energetic materials to thermally degrade agents via rapid heating. For example, the exothermic reaction of composite metal powders has previously shown promise in neutralizing BWAs, thus making them promising candidates for neutralizing CWAs.[4] Here, we describe the development and methodology behind leveraging two measurement systems to observe the neutralization of a CWA simulant via combusting composite reactive metal powders while monitoring background gas temperatures. Composite reactive metal powders exhibit a wide variety of ignition and combustion characteristics that can be controlled by tuning their chemistry and morphology.[5–9] To understand their interaction with CWA simulants, two decomposition pathways must be considered: (a) homogeneous thermal defeat due to the elevated temperatures produced by combusting the metal particles and (b) local thermal and chemical defeat of CWA simulants that are directly adjacent to the combusting metal powders. Combustion of the metal powders will result in high local temperatures[10–12] and oxides that are known to neutralize CWAs and their simulants.[13–24] Given some metals combust in the gas phase while others combust in a condensed phase, the size and number density of the metal oxides in a combustion plume can vary tremendously with chemistry. Metals such as Al and Mg that burn in the vapor phase will produce a high density of nanoscale oxides, which can be seen as soot or smoke in a combustion plume. In contrast, metals such as Ti or Zr burn in the condensed phase, producing a lower density of microscale oxide particles. Thus, even though their local heat production may be equivalent, the contribution of their oxide products to simulant neutralization may be very different. Therefore, to optimize metal powders for simulant neutralization, diagnostic systems are needed to monitor both temperature and CWA molecular evolution at fast timescales (≥100 Hz) and over short durations (<5 s). The decomposition of CWAs and CWA simulants, such as diisopropyl methylphosphonate (DIMP), dimethyl methylphosphonate (DMMP), dimethyl nitrophenyl phosphate (DMNP), and triethyl phosphate (TEP), has been studied using both experimental and computational methods,[13–15,17–33] yet the decomposition pathways are complex and are sensitive to experimental design and measurement equipment. Zegers and Fisher *et al.*[25] identified a decomposition pathway for DIMP using a combination of Gas Chromatography-Mass Spectrometry (GC-MS) and Fourier transform infrared spectroscopy (FTIR),

$$\begin{array}{c} DIMP \longrightarrow IMP + Propene \longrightarrow MPA + Propene \\ \downarrow \\ MOPO + 2 - Propanol \end{array} \quad . \tag{1}$$

While the GC-MS technique has high sensitivity to changes in the molecular structure, the technique requires live analyte sampling and has limited capability in measuring fast phenomena.[16,25–27] Absorption spectroscopy techniques, such as Fourier transform infrared spectroscopy (FTIR), are popular alternatives for monitoring organic compounds, such as DIMP, TEP, and DMMP,[14–16,34,35] as their functional groups absorb at different frequencies and absorption is linearly related to concentration, as shown by the Beer–Lambert law as follows:

$$\log\left(\frac{I_o}{I}\right) = A = \varepsilon \cdot l \cdot c, \tag{2}$$

where $A$ is the absorbance, $I$ is the transmission intensity, $I_o$ is the background intensity, $\varepsilon$ is the molar absorptivity coefficient, $l$ is the path length, and $c$ is the concentration of measured species. Another advantage of absorption spectroscopy is the ability to monitor experiments remotely without requiring live analyte sampling or post-reaction material analysis.[13,16,25,26,29] FTIR provides high wavelength resolution and sampling over a wide wavelength range and, thus, can be used to monitor multiple functional groups within the IR spectrum. This provides the ability to accurately monitor the evolution of initial, intermediate, and final by-products. However, probing a broad wavelength range at high resolution requires time resolutions greater than one Hz, a similar drawback to MS.

Quantum Cascade Lasers (QCLs) have also been used for absorption spectroscopy to monitor organo-phosphates at repetition rates well above kHz and approaching MHz (8–9). The drawback to QCLs, however, is a very narrow cross section (0.1 $cm^{-1}$) that can only probe single functional groups. This hinders monitoring of both simulants and their by-products, particularly when CWA simulants and their by-products have overlapping peaks in the IR regime.

External cavity quantum cascade lasers (ECQCLs) are an emerging technology capable of operating at scan rates of a few hundred Hz, high resolution, and moderate spectral range. They have been used to monitor organic compounds in chemical environments but come with a high degree of development complexity. The ability to design, build, and tailor these systems for unique experiments is difficult without qualified expertise.[36,37]

An exciting alternative to the above methods is a rapid scanning dispersive spectrometer with kHz repetition rates and the ability to probe moderate wavelength ranges with relatively simple optical designs.[38–41] These systems are low cost and highly adaptable to meet specific experimental needs. A two-slit system with a rotating grating was recently developed and used to characterize the neutralization of DIMP.[42] Our spectrometer design builds from this earlier success but employs a rotating polygonal mirror instead of a rotating grating to ease cost and system complexity. We refer to the new tool as a Polygonal Rotating Mirror InfraRed Spectrometer (PRiMIRS), and we calibrate it using three compounds: anisole, ethyl benzoate, and $NH_3$. We then use this tool to characterize the rapid decomposition of DIMP in the presence of dispersed combusting metal powders.

PRiMIRS does have limitations in the total spectral bandwidth compared to FTIR systems, and IR spectroscopy, in general, suffers from the inherent problem of signal overlap when similar functional groups are present in different species. To mitigate these drawbacks and be able to analyze decomposition, PRiMIRS was designed to probe a region with the following requirements:

1. Multiple distinct features of DIMP must be present within the wavelength range.
2. Distinct features of one or more by-products must be within the wavelength range.
3. Locations within the probed region must exist where signals corresponding to DIMP are low and any signal enhancements would be indicative of by-product formation.

In addition to measuring DIMP, it is necessary to monitor properties unique to the combusting metal particles to understand their contributions. One such property is the rise in the surrounding gas temperature as particles combust. Thermal measurements are largely performed using two primary techniques, either using thermocouples or leveraging optical techniques. While thermocouples demonstrate a high level of accuracy and precision, their response times are inherently slow, especially if there are significant changes to the environment, approaching Hz or greater.[43–46] Given that we are remotely monitoring DIMP on the order of 100 Hz timescales, thermocouples can be challenging for this application.

Optical measurement techniques include infrared (IR) cameras as well as spectroscopic techniques. IR cameras leverage IR radiation typically from surfaces that are calibrated for temperatures and, thus, are not well suited for measuring changes in gas temperature within a sample cell. Emission spectroscopy is used in combustion science and is highly successful at reporting high temperatures associated directly with burning species. Temperature can be calculated from recorded emission intensities, as shown in Ref. 47. Examples of temperature measurements obtained through two and three-color pyrometry techniques are shown in Refs. 48–51. These types of measurements are more suitable for understanding the local high temperatures associated with or very near the burning metal species. Our interest is specifically toward the contribution of burning metal particles to rises in background or global gas temperatures, which is more appropriate to the spatial distribution of DIMP vapor. In order to obtain these measurements, we leverage absorption spectroscopy as the ambient gas itself is not an emitting species.

Tunable diode laser absorption spectroscopy (TDLAS) is the specific technique we employ and has been demonstrated previously by Mattison *et al.*[52,53] and followed up by Murzyn *et al.*[54] Development of new experimental systems, techniques, and validation of their measurements is always a challenging endeavor. Alternative to the specific applications described in this study, additional publications by Kasmaiee *et al.* describing system development and their validation through experimental and computational methods can be reviewed.[55–57]

This study presents the first known measurements for prompt defeat of CWAS using combusting composite metal powders without the use of high explosive (HE) as an initiator. The necessity to improve efforts toward the prompt defeat of CWAS was described earlier in addition to individually attractive properties of combusting metal particles. For a comprehensive analysis, we monitor the small but rapid rise in background gas temperatures due to the combusting metal powders using a TDLAS system centered just below 1390 nm. This range was selected to extract temperatures from measured $H_2O$ spectra collected at 100 Hz, matching the operating rate of PRiMIRS. Here, we describe the PRiMIRS and TDLAS systems and demonstrate that these systems can be used in conjunction to study the prompt defeat of CWA simulants using combusting metal powders. Devising the first approach to evaluating composite metal powder performance on CWAS neutralization from which future work can be built is the true innovation of this study.

## II. MATERIALS AND METHODS

### A. Experimental cell (DIMP cell + vapor cell)

The experimental cell in Fig. 1 was constructed from a stainless steel (SS) reducing tee with a length of 5.12″, KF40 end ports, and a KF16 transverse port (Kurt J Lesker QF40X16T). ZnSe viewports (Kurt J Lesker QF40-150-VPUZC) are attached to the ends to allow the passage of IR light to measure DIMP with PRiMIRS, and sapphire glass viewports (Accu-glass Products 112 669) are used to measure global gas temperatures using TDLAS. A KF16 to 1/8″ NPT Swagelok needle port allows for sample loading using a syringe. A KF16 thermocouple feedthrough (Kurt J Lesker EFT0021038Z) was used in place of the needle port for TDLAS calibrations. The SS sample cell is wrapped in heating tape (BriskHeat BS0051020L), insulated with ceramic fiber, and housed inside an acrylic box to minimize thermal gradients within the cell. A metal substrate with dimensions of $5 \times 5 \times 0.01$ cm$^3$ is placed under the loading zone to normalize the thermal mass that liquid DIMP would contact between experiments and to cover a second transverse port required for electrical feedthroughs. A stainless steel (SS) substrate is used for these experiments to match the material forming the cell walls. The cell is preheated externally to temperatures between 115 and 230 °C and held at the chosen temperature for 45 minutes to allow thermal gradients to equilibrate within the cell. Background PRiMIRS measurements are taken prior to sample introduction in order to calculate absorption spectra from $(I/I_o)$. After preheating the cell, 20 $\mu$l of the liquid sample is loaded into the cell through the needle port using a syringe. This generates adequate vapor for spectroscopy without saturating the detector. Picoscope 6 waveform software is used to record spectra collected from the oscilloscope (Picoscope 5443D). The function generator operating the polygonal mirror sends a second signal to trigger the oscilloscope and sync data acquisition to the repetition rate provided by the polygonal mirror.

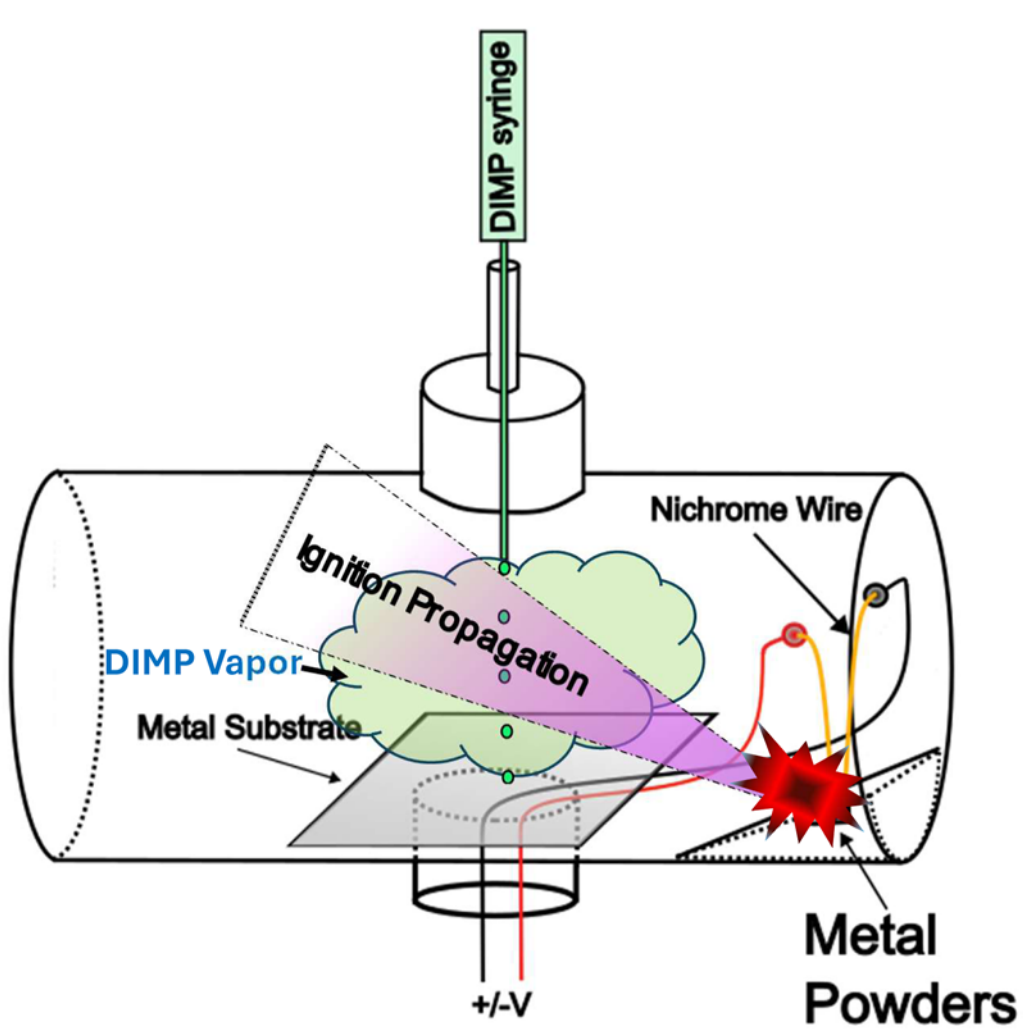

**FIG. 1.** Stainless steel experimental cell. Metal powders are loaded on the bottom right, are ignited using a resistively heated nichrome wire, and then propagate across the cell while combusting. During DIMP experiments, the cell is preheated, liquid DIMP is loaded through a needle port onto a stainless steel metal substrate, DIMP vapor is generated, and powders are ignited.

For data acquisition within the Picoscope software, two particular measurement parameters need to be adjusted to capture the desired amount of information at the required time scales. These parameters include the timebase, which determines what length of time is actively captured in each waveform, and the number of samples, which determines how many individual data points were collected in that captured waveform. These two parameters do not have infinite flexibility as there will be a co-dependence depending on how far each parameter is stressed. The specific settings chosen for the measurements made in this study are described here:

1. The slow capture parameter consisted of averaging raw spectral measurements collected over 3 s at a 1.3 kHz capture rate (~4000 measurements). Within the Picoscope software, the number of samples is set to 10 000 and the timebase is 500 $\mu$s.
2. The fast capture parameter averaged 13 spectral measurements with a raw capture rate of 1.3 kHz, resulting in 100 Hz temporal resolution with signal to noise ratio (SNR) enhancement through spectral averaging. Picoscope settings were adjusted to the following parameters: 500 waveforms were collected with rapid pulse technology ($\leq\mu$s delay between waveforms); each waveform contained 50 000 samples, and the timebase was set to 10 ms.

For the ignition of metal powder, a similar cell was constructed from a reducing cross with KF40 end ports and two transverse KF16 reducing ports perpendicular to the KF40 tube. An electrode feedthrough is used to run current through a resistively heated nichrome wire that ignites powder held in a machined SS mount placed at one end of the cell. The feedthrough and wires are covered by the SS substrate to avoid contact with liquid DIMP. When the DIMP vapor signal reaches a relative maximum (>2 min), a Matlab program begins data acquisition and after a brief delay enables the power supply to resistively heat the wire and ignite the metal powders. (Al–8Mg):Zr composite metal powders[6] were used for the experiments performed in this study and prepared using a Retsch planetary mill. An equal atomic percent of Al–8Mg and Zr starting powders were milled for 45 min at a ball to powder ratio of 3 using hexane as the process agent and then sieved below 75 $\mu$m. More details on this technique can be found in Ref. 6.

Vapor temperature profiles were first measured as a function of time in an identical cell without contamination from DIMP in order to establish an appropriate powder loading and ignition protocol. 125 mg of (Al–8Mg):Zr powder provided reliable ignition, a clear rise in vapor temperature, and a combustion event that was not too dense to be probed by TDLAS. Several repetitions were performed to gauge rises in global gas temperature starting at both room temperature and 115 °C in preparation for DIMP experiments. Subsequently, the experiment was then performed with the same protocol in a cell including DIMP vapor that was monitored using PRiMIRS during the metal powder combustion.

## B. Polygonal rotating mirror infrared spectrometer (PRiMIRS)

The PRiMIRS 2-slit spectrometer Fig. 2 utilizes a polygonal rotating mirror to deliver IR light to a stationary grating while varying the angle of incident light as a function of time. An IR light source (Boston Electronics IR-Si207) is operated at 12 V and 1.62 A to reach the requisite thermal temperature of 1260 °C for the production of broadband IR light. A 100 mm ZnSe lens sits at the focal length in front of the light source to collimate broadband light through the sample cell (detailed later in Fig. 3). The collimated light is directed through an entrance slit set to 0.025″, after which a 75 mm ZnSe lens system focuses light after the slit to reflect off a rotating polygonal mirror (Precision laser scanning Gecko-33-OTS). Another 75 mm ZnSe lens recollimates reflected light toward a system of parabolic mirrors. Gold coated 60° off-axis parabolic mirrors with 4″ focal lengths are positioned to diffract light off an IR grating with a 10 $\mu$m blaze wavelength and a groove density of 75 gr/mm. Finally, the diffracted beam passes through a 6.0 $\mu$m long pass filter placed before an exit slit set to 0.025″ for signal collection with a liquid $N_2$ cooled detector (Infrared Associates MCT-12-1.0). The detector is primed for a 10 $\mu$m signal peak and operates using a pre-amplifier with 150 kHz electrical bandwidth (Infrared Associates MCT-1000). Its wavelength range can be modified by adjusting grating positioning with respect to the optics, and the wavelength resolution can be tuned by modifying the groove density, the blaze angle of the grating, or the slit openings. The grating is positioned to provide a wavelength range of 7.4–9.6 $\mu$m and a wavenumber resolution of 15 $cm^{-1}$, which are appropriate for probing distinct features of the CWA simulant of interest, DIMP, and potential by-products, in the fingerprint region of the IR spectrum. Prior to experimental runs, spectra are recorded with 8.0, 8.5, and 9.0 $\mu$m filters (Thorlabs FB8000-500, FB8500-500, FB9000-500). Using vendor provided spectral profiles of these filters, a linear fit is calculated with the measured spectra to determine the wavelength range with $R^2$ values exceeding 0.98.

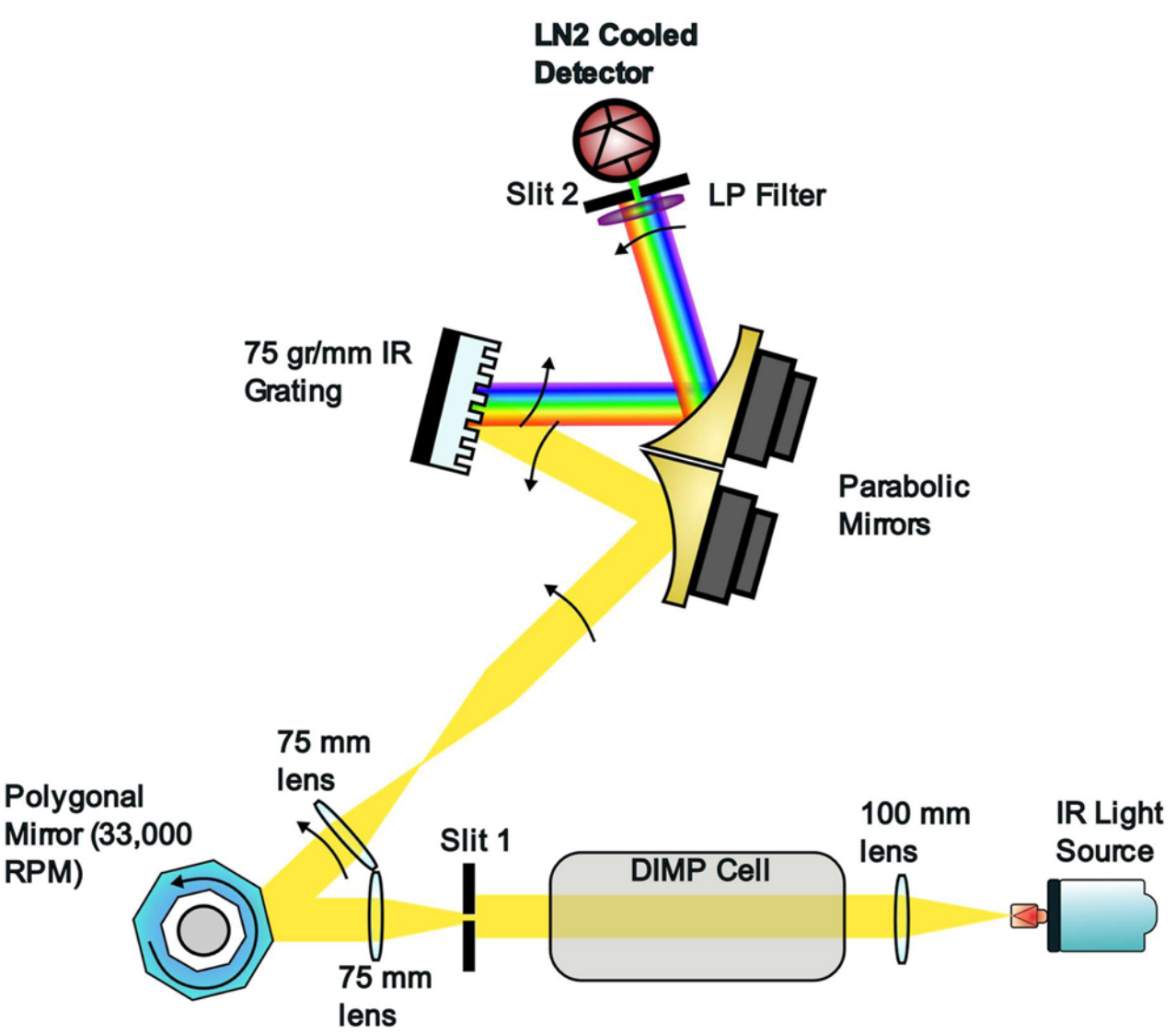

**FIG. 2.** Schematic diagram of the PRiMIRS spectrometer developed to monitor IR spectra of DIMP at 100 Hz or greater to measure the efficacy of combusting metal powders for CWAs.

## C. Tunable diode laser absorption spectrometer (TDLAS)

The TDLAS system [Fig. 3(a)] was constructed to measure gas temperatures in the presence of combusting metal powders. A Thorlabs combined laser driver/controller (CLD1015) is used to control a laser diode centered at 1390 nm (Eblanaphotonics EP1392-DM-B).

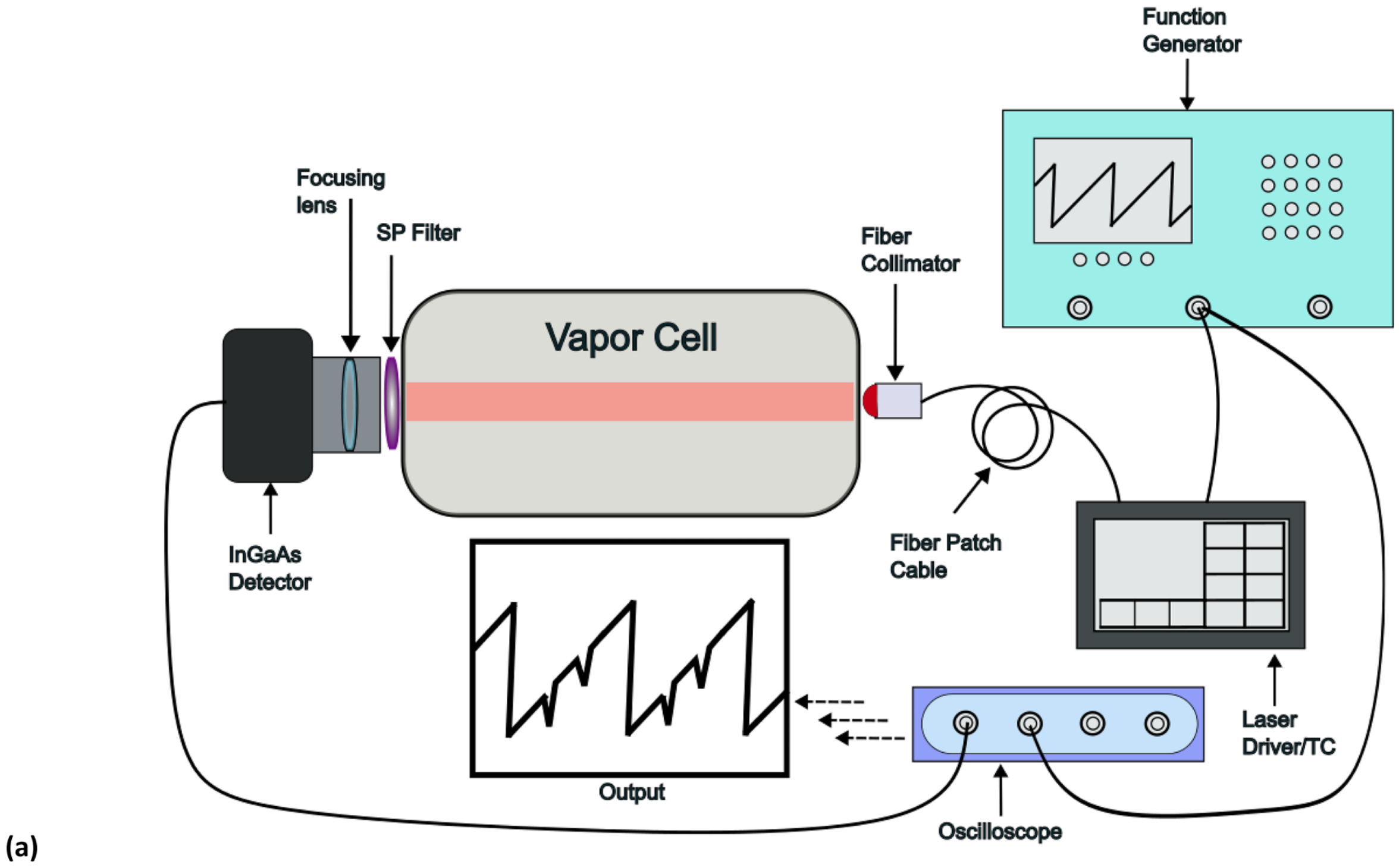

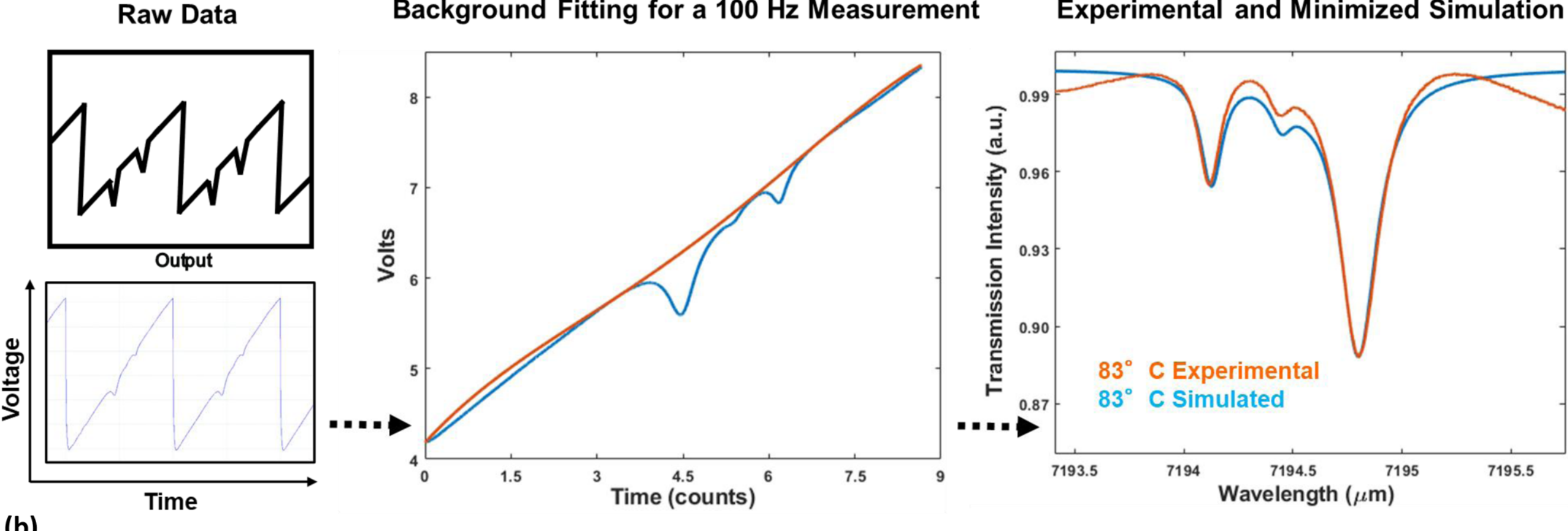

**FIG. 3.** (a) Schematic of the TDLAS system developed to monitor background gas temperatures during the combustion of metal powders. (b) A Matlab minimization algorithm was adapted to extract vapor temperature by comparing measured transmission spectra to simulated spectra with parameters available on the HITRAN database.

A function generator (Tektronix AFG31051) feeds a waveform to the laser driver to modulate the laser, resulting in a scanning range of ~0.6 nm at 100 Hz. The laser driver/controller was set to operate at 14.00 °C, and the current was centered at 73.0 mA. The laser is collimated through the experimental cell (Thorlabs F280APC-C), passes through a bandpass filter (Edmund Optics 85 900) placed directly outside the cell windows, and focuses into an InGaAs detector (Thorlabs DET10C2). The signal is collected using a Picoscope 5443D oscilloscope. A BK precision power supply is set up as a serial communication device to allow for external triggering and sequential data acquisition into a PC operating Picoscope 6 software. Temperature is extracted using a Matlab program developed in Ref. 54 and adapted to fit our experimental parameters. This algorithm (1) generates experimental transmission spectra from raw voltage measurements, (2) simulates spectra as a function of temperature, pressure, and concentration using parameters available on the HITRAN database,[58–64] and then (3) utilizes a minimization algorithm between experimental and simulated spectra to report temperature, illustrated in Fig. 3(b). The system was calibrated by tuning the algorithm to match steady state temperatures that were validated using an internal thermocouple.

## III. RESULTS

### A. Evaluation of PRiMIRS, a spectroscopic system developed to monitor DIMP at fast time scales

The performance of the PRiMIRS spectrometer was evaluated using three compounds: anisole (Alfa Aesar < 100%), ethyl benzoate (Millipore Sigma 99+%), and $NH_3$ (Alfa Aesar, ammonium hydroxide, 28% $NH_3$. The wavelength range was calculated using 8.0, 8.5, and 9 $\mu$m filters with 0.5 $\mu$m bandwidths prior to each measurement. Spectral measurements were taken at room temperature under steady state conditions with a repetition rate of 1.3 kHz and the slow capture parameters within the Picoscope software. These capture parameters were appropriate for monitoring compounds at conditions where rapid spectral transformations (<1 s) were not expected to occur, resulting in a very high SNR. Measured spectra for ethyl benzoate and anisole are illustrated in Fig. 4(a) with spectral overlays obtained from the NIST database.[65,66] These measurements demonstrate the ability of PRiMIRS to appropriately distinguish relevant spectral features within these compounds.

Simulated NH3 spectra were generated using parameters available in the HITRAN database.[58–64] A Gaussian convolution was applied to these spectra to simulate spectral profiles with FWHM resolutions ranging from 14 to 17 $cm^{-1}$. A comparison between simulated and experimental spectra is displayed in Fig. 4(b) and demonstrates that the spectral resolution of PRiMIRS is better than 15 $cm^{-1}$. In addition to the wavelength calibration described in Sec. II B using 8.0, 8.5, and 9.0 $\mu$m filters, linear fits are calculated using reference data for anisole and ethyl benzoate and simulated data for NH3. A linear wavelength model presents with $R^2$ and adjusted $R^2$ values exceeding 0.99 for all compounds, and the process is demonstrated in Appendix Subsection 1.

Low throughput measurements on DIMP (Alfa Aesar 95%) were performed under isothermal conditions at temperatures between 115 and 185 °C using slow capture parameters. Spectra taken at 115 °C over a 3 h period are shown in Fig. 5(a). A steady reduction in signal intensity is observed over time; however, there are no indications of decomposition products as signal enhancement is not present in any region, and variations in the relative intensity of the peaks within the DIMP spectra are not observed. Since the boiling point of DIMP is 215 °C, condensation outside of the optical path is a likely contributor to signal reduction over time at these timescales as the viewports and loading port flanges are cooler than the bulk of the chamber. Because of these reductions in signal over time, DIMP decomposition measurements due to powder combustion were restricted to several seconds. Higher throughput characterizations of DIMP spectra were also performed using the previously defined fast capture parameters and results are displayed in Fig. 5(b). While the SNR was lower at faster collection times, the SNR is still satisfactory while meeting the experimental demands in speed. The primary features of DIMP (two peaks below 8 $\mu$m and three peaks above 8 $\mu$m) are still sufficiently resolved such that they can be monitored in the case where metal powders combust in the chamber at faster timescales. In this case, with no combusting particles, there is no evidence of DIMP decomposition over very short times.

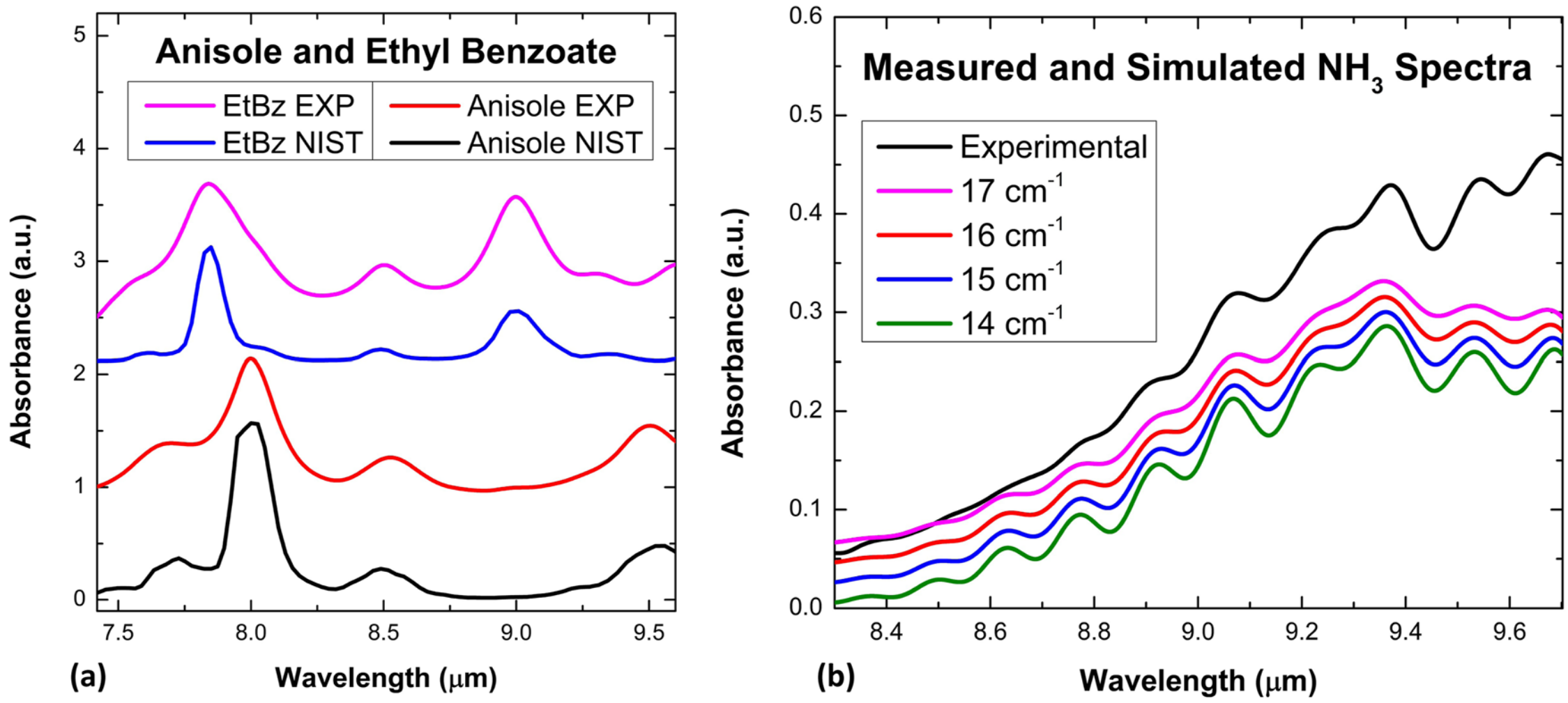

**FIG. 4.** (a) Measured absorption spectra using PRiMIRS and reference spectra from the NIST database are shown for anisole and ethyl benzoate. (b) Measured absorption spectra for NH3 using PRiMIRS and simulated spectra using parameters available on the HITRAN database are shown at various wavenumber resolutions. Spectra in (a) and (b) are vertically offset for clarity.

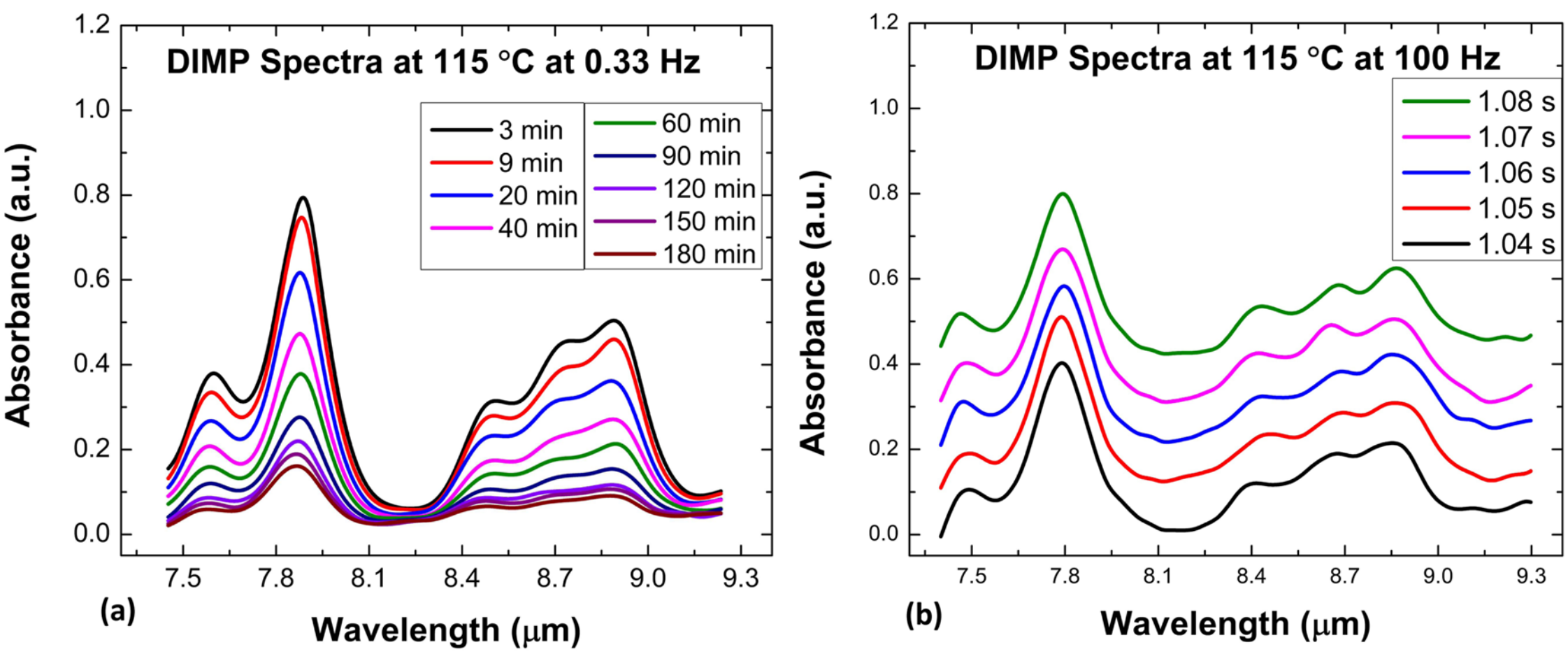

**FIG. 5.** (a) DIMP absorption spectra obtained using the slow capture parameters are shown for select time points at 115 °C over 180 min. Measured absorbance decreases over time. (b) Sequential absorption spectra of DIMP obtained using fast capture parameters at 115 °C are vertically offset for clarity.

### B. TDLAS to measure global gas temperature contributions of composite metal particles

A TDLAS system was developed in order to measure gas temperature contributions from burning metal particles. As depicted earlier in Fig. 3(b), each data point using TDLAS was generated by minimizing the mismatch between simulated and measured spectra. The previously described algorithm sequentially processes these spectra, providing a temperature profile much faster than can be obtained by manual fitting.

The TDLAS measurements were compared with alternative measurements made from a thermocouple placed in the center of the cell that was held at room temperature, 115, and 150 °C. There is a strong correlation between the TDLAS and thermocouple measurements in Fig. 6(b). The temporal resolution of the TDLAS system was set to 100 Hz, matching the capability of PRiMIRS. While temperature spectra can be simulated over a broad range of temperatures and at increments far below 1 °C, the variance among spectra for small changes in temperature is minimal. Experimental spectra will also contain noise where the simulated spectra do not. Using the minimization algorithm with a set number of iterations, the RMSE of TDLAS temperature measurements is evaluated to be 2.8 °C at room temperature, 2.3 °C at 115 °C, and 5.1 °C at 150 °C. The data acquisition speed (100 Hz) provide enough data points for a reasonable temperature profile, given the application.

125 mg of ball milled (Al–8Mg):Zr powders were loaded into an SS mount at one end of the cell and ignited using a resistively heated nichrome wire at room temperature. The orientation of the mount allows the ignited particles to propagate across the tube toward the other end. After ignition, the temperature quickly rose to a maximum (<0.25 s) and then relaxed to a moderately elevated temperature in 0.5 s, cooling completely over several additional seconds. Several measurements were performed, demonstrating temperature increases of 103.3 ± 42.1 °C, n = 3; a profile is shown in Fig. 6(a). A thermal ramp at this temporal scale is difficult to capture using conventional measurement tools. Fast thermocouples are available; however, the ability to remotely probe is a strength of TDLAS, and for future applications, TDLAS has demonstrated operation at much faster timescales.[54] The experiment was also performed while preheating the cell to 115 °C prior to ignition [Fig. 6(c)] as these conditions are used during DIMP experiments. Temperatures reported from path averaged measurements within the preheated cell were still measured to be 100.7 ± 14.0 °C, n = 3, while the peak temperature remained below 230 °C. The rise and fall in temperature are demonstrated with ~50 data points preceding and following the peak temperature. It is important to consider that if there is variation in temperature along the measurement path, the measured spectra will be affected by these regions of fluctuating temperature. However, our goal in this study was to present the means to provide a thermal parameter to differentiate combustion events, more so than to report the absolute temperature of associated events.

### C. Decomposition of DIMP evaluated using PRiMIRS

The operating wavelength range of 7.4–9.6 $\mu$m for the PRiMIRS system was selected to probe for two primary decomposition products from Eq. (1), isopropyl methyl phosphonate (IMP) (Synquest Laboratories) and isopropyl alcohol (IPA) (Fisher Scientific). Stable forms of these by-products were obtained to enable in-house reference measurements. Steady state spectra were collected separately for DIMP, IMP, and IPA with the cell preheated to 115 °C, and the resulting spectra are shown together in Fig. 7(a). Three regions of particular interest are identified. Region 1 (7.4–7.9 $\mu$m) contains a strong signal from DIMP but weaker signals from by-products IMP and IPA. Region 2 (7.9–8.3 $\mu$m) and region 3 (9.0–9.3 $\mu$m) correspond to areas where the contribution of DIMP is low, but the signals for IMP and IPA are strong. If decomposition were to occur, the expected spectral transformations would be as follows: the signal in region 1 should drop, while the signal in regions 2 and 3 should rise, indicating a conversion of DIMP to by-products, such as IMP and IPA. Mathematical combinations of DIMP and by-products

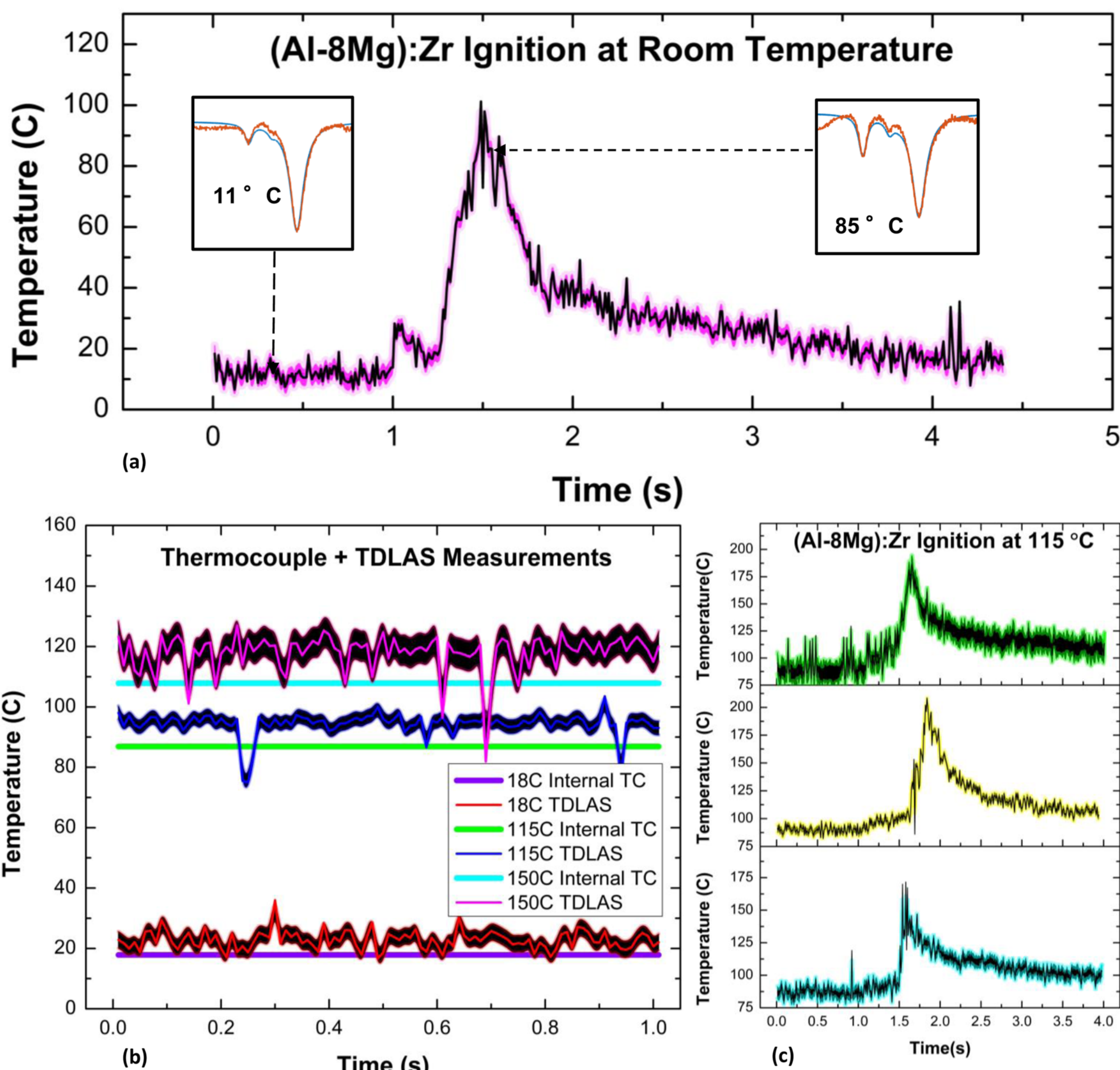

**FIG. 6.** (a) Thermal profile during the ignition and combustion of 125 mg of (Al–8Mg):Zr powders at room temperature. Temperatures were determined by minimizing the differences between measured and simulated spectra as shown in the insets for two different time points. Additional minimizations between experimental and simulated spectra are available in Appendix Subsection 2. (b) A thermocouple was placed inside the SS cell in conjunction with TDLAS at different steady state temperatures, and measurements are reported. (c) Thermal profiles during the ignition and combustion of 125 mg of (Al–8Mg):Zr powders after the SS steel cell is externally preheated to 115 °C. Associated error was determined from the standard deviation of the stable pre-fire regime and can be viewed in Appendix Subsection 3.

IMP and IPA are shown in Fig. 7(b). Absorption spectra are additive, and this plot shows what the spectral evolution is expected to look like as DIMP converts to by-products IMP and IPA. Of course, there are additional by-products that have a spectral contribution; however, we cannot make reference measurements for all of them as they are not readily available.

Experiments to explore homogeneous thermal decomposition were first performed by preheating the DIMP cell to 230 °C and establishing steady state conditions. 20 $\mu$l of DIMP was then loaded into the cell to generate vapor, and rapid measurements were taken at 100 Hz using fast capture parameters. In Fig. 8(a), neither signal enhancement nor reduction is seen in regions 1, 2, or 3, suggesting that decomposition of DIMP does not occur in a 5 s timeframe in homogeneous thermal environments up to 230 °C. A heterogeneous experiment was then performed where the experimental cell was preheated to 115 °C, 20 $\mu$l of DIMP were loaded to generate DIMP vapor, then 125 mg of (Al–8Mg):Zr powder was ignited, and the combusting powders propagated across the DIMP vapor. In these heterogeneous experiments, we observed a signal decrease in region 1 and signal increases in regions 2 and 3 over two seconds [Fig. 8(b)]. These spectral transformations are strong indicators of DIMP decomposition and by-product formation.

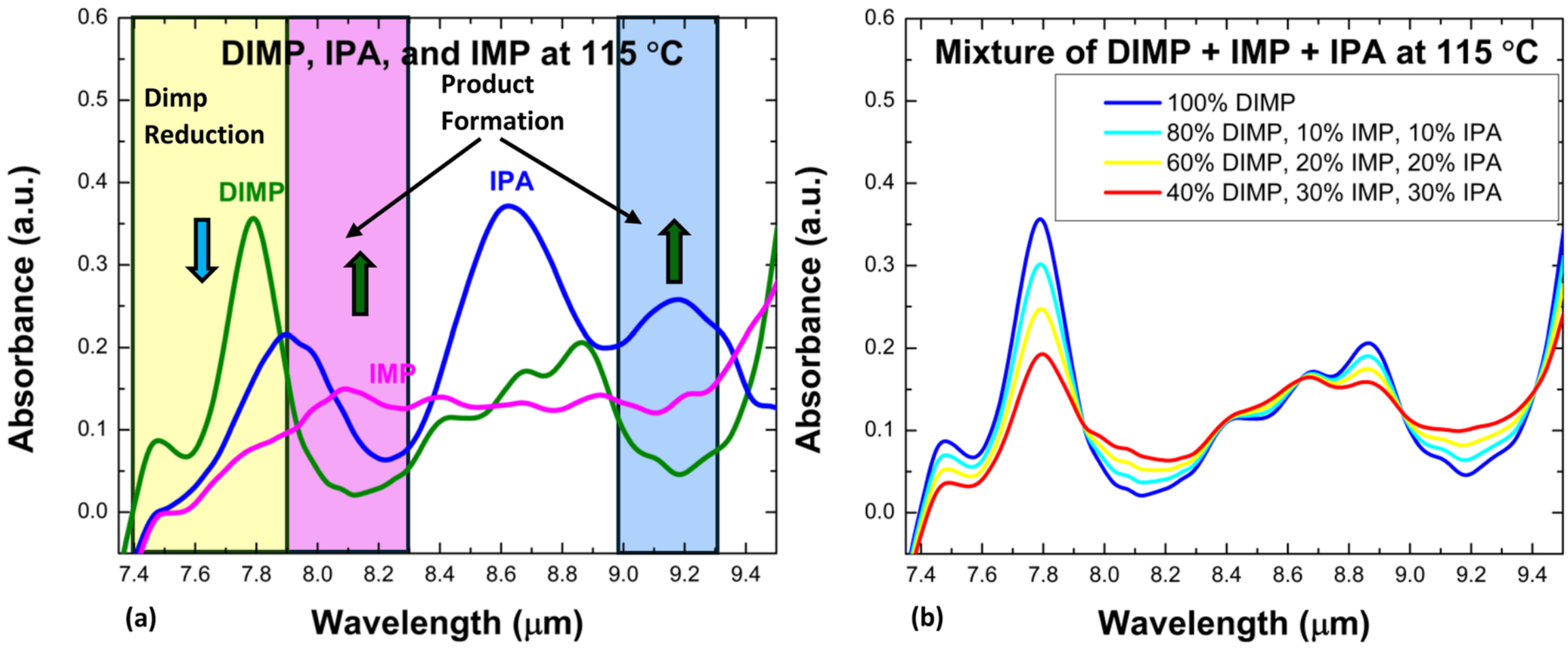

**FIG. 7.** (a) Steady state absorption measurements for DIMP and primary decomposition products IMP and IPA. Measurements for each gas were taken individually with the SS cell externally preheated to 115 °C. (b) Mathematical combinations of components from (a) scaled to various concentration levels for qualitative analysis.

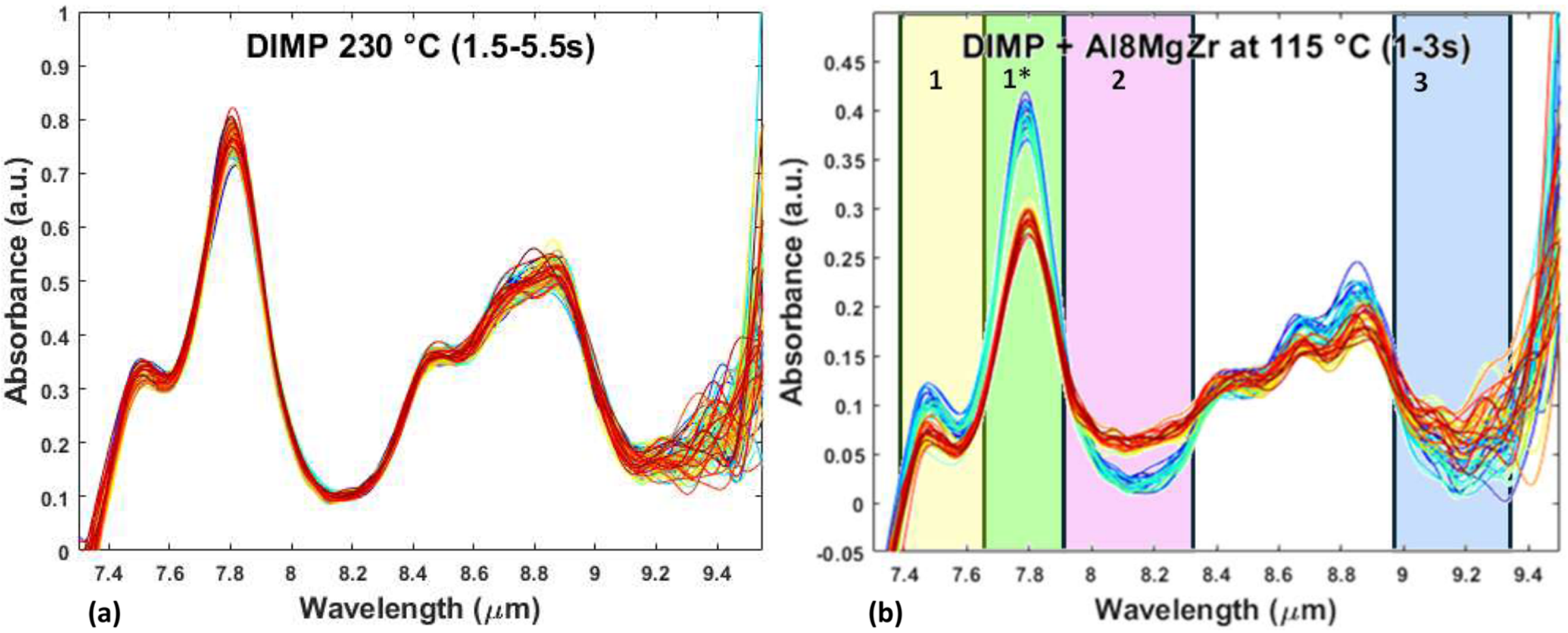

**FIG. 8.** (a) Measurement of absorption spectra of DIMP within the SS cell that is preheated to 230 °C to establish a homogeneous thermal environment. (b) DIMP is loaded in a SS cell preheated to 115 °C, (Al–8Mg):Zr powders are ignited, and absorption spectra are recorded. Select time points between 1 and 3 s are plotted with a thermal color palette corresponding to time. Early time points are blue shifted and later time points are red shifted. Spectra from the middle time points were removed to highlight the change between earlier and later times. Region of the spectra are numbered for further analyses.

Subsequent analysis separates region 1 and region 1*. This is to account for possible by-product formation of species on which we cannot make our own reference measurements and is further explained in Sec. IV C.

In Fig. 8(b), the earlier times are blue shifted, the later times are red shifted, and the middle times are removed to clarify the spectral changes over time. To comparatively quantify the extent of decomposition and product formation, the peak areas are integrated from the 100 Hz collected spectra in regions 1, 1*, 2, and 3 and sequentially plotted in Figs. 9(a)–9(d), respectively. The integrated intensities in regions 1 and 1*, which correspond to DIMP, decrease, while the integrated intensities in regions 2 and 3, which correspond to decomposition products IMP and IPA, increase post-combustion. The sudden drop in integrated absorbance around 1.5 s is due to the combusting metal powders releasing infrared radiation across mid and long wavelength regions, which cannot be accounted for in our background measurements.[67] These brief timeframes during which we cannot account for background signal will not be considered in future quantitative analyses.

## IV. DISCUSSION

### A. PRiMIRS

PRiMIRS: The PRiMIRS system developed here is a true bench top spectrometer; all components are available off the shelf and no machining or non-standard power requirements are necessary to build and operate this system. The comparison of experimental spectra to reference spectra (Anisole and Ethyl Benzoate) that are available on the NIST database, as well as simulated spectra using parameters from the HITRAN database for NH3, demonstrated the ability of PRiMIRS to separate distinct spectral features with a FWHM spectral resolution better than 15 $cm^{-1}$. The polygonal rotating mirror specifications allow for raw collection rates up to 4.4 kHz.

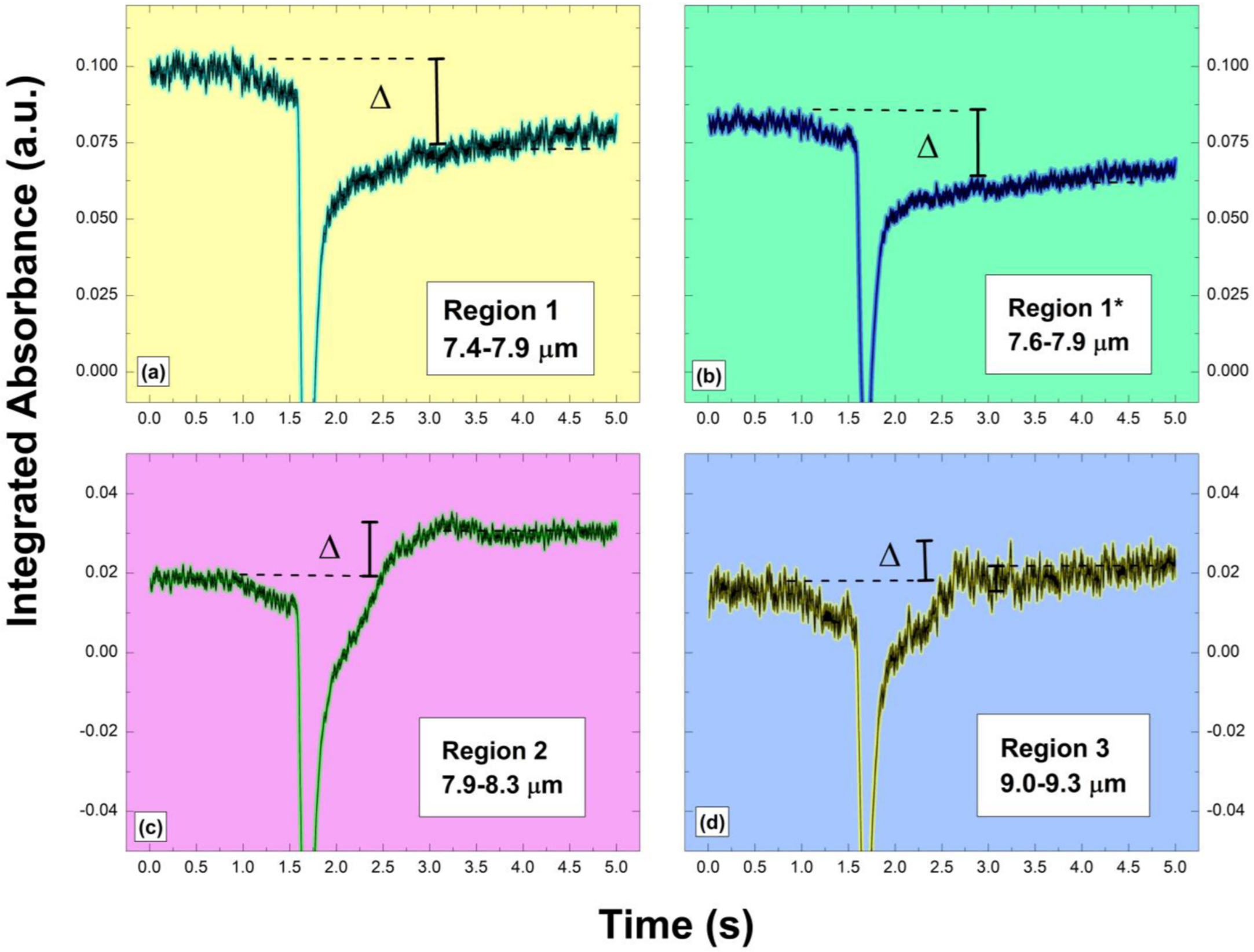

**FIG. 9.** (a)–(d) Integrated absorbance spectra for DIMP acquired at 100 Hz are shown for corresponding regions previously identified in Fig. 8(b). Associated error was determined from the standard deviation of the stable pre-fire regime and can be viewed in Appendix Subsection 3.

While PRiMIRS is an appropriate system to monitor DIMP decomposition at sub-second timescales, the system does present some limitations. PRiMIRS does not have the overall wavelength scanning range that can be probed using systems such as FTIR. This means that the customized wavelength range of PRiMIRS must be tuned to selectively measure transitions of interest. While we cannot measure unique spectral signatures of all possible decomposition products, we can capture a distinct spectral transformation that accounts for signals unique to DIMP and separate from some of its primary decomposition products. In addition, obtaining spectral measurements at fast rates (>kHz) lowers SNR especially since the intensity of broadband light at LWIR regimes is low. The low intensity light is further reduced as it passes through each component of the optical system (viewports, lenses, mirrors, and slits). To overcome this limitation for the given experiments, waveform averaging was leveraged within the Picoscope software to reduce noise and generate spectra with a very high SNR at <Hz and a reasonably high SNR at 100 Hz, as shown in Figs. 5(a) and 5(b), respectively. While we utilized 100 Hz in the experiments performed in this study, the flexibility in operational speed and tunability of components within PRiMIRS offers a unique system that can be tailored to a variety of experiments.

We also note that optical aberrations build as light passes through each component in an optical system. This can affect wavelength linearity, particularly at the bounds of the wavelength range.[68] Since this was a custom-built spectrometer, it was important to take our own reference measurements for compounds of interest as spectral signatures may not line up exactly with the robust FTIR measurements that are available in spectral databases. Grating rotation and alignment were adjusted such that the wavelengths of interest for data analysis were comfortably within the bounds of the instrument's capability. For this reason, we do not draw conclusions from data measured below 7.4 μm and above 9.3 μm during our DIMP experiments. Furthermore, low intensity broadband light is an obstacle to differentiating weak features in addition to previously mentioned SNR limitations. For example, Fig. 5(b) demonstrates the poor differentiation of ammonia features below 8.8 $\mu$m that present with very low intensities in the simulated reference spectra. While performing waveform averaging at faster collection rates may help reduce background noise, this technique can also obscure weaker spectral features. Absorption spectroscopy is also a path averaged measurement technique. This can be problematic if the application requires pristine spectral accuracy since variations in species concentration and temperature along the measurement line will have contributions to the spectra. However, the main objective of this study was to produce a measurement system to remotely probe DIMP during a combustion event with sufficient resolution and high speed to observe the rapid decomposition of DIMP. These objectives were met by PRiMIRS.

## B. TDLAS

TDLAS: The TDLAS system was employed to monitor global gas temperatures generated by combusting metal particles to assess their thermal contributions during DIMP experiments. Measurements were performed at 100 Hz matching the acquisition rate of PRiMIRS, and simultaneous measurements were taken using a thermocouple at steady state temperatures in Fig. 6(b). We do note a lower reported temperature from the internal thermocouple as opposed to the spectral measurement. Reasons for this discrepancy could be the following: When using thermocouples to measure gas temperatures, the length of the high temperature region surrounding the thermocouple wire has been shown to impact the reported temperature.[69] Additionally, thermocouple operation is dictated by the Seebeck effect, where voltages vary between dissimilar metals at different temperatures.[70]

This phenomenon occurs in a circuit between the hot and cold junction of a thermocouple where the cold junction is typically a static reference temperature. Due to experimental constraints between our sample cell and the thermocouples to which we had access, the cold junction of our thermocouple used during experimentation could not be isolated in a cold bath and had to remain near the end of the heated cell. Thus, reported temperatures at the hot junction will be lower than the actual temperature. Expectedly, as the cell temperature increases, this delta will also increase. This can explain the higher reported temperatures measured using TDLAS since heat transfer phenomena do not affect optical analysis. More importantly, the calculated temperature from spectral analysis was consistent over a duration of several seconds at 100 Hz acquisition rates. This provides confidence going forward that if there are significant relative changes in thermal profiles between different metal powder formulations, they can be detected using TDLAS.

While thermocouples had some challenges in this particular application, they are a well-known thermal measurement tool and standard to a variety of applications. In this case, the goals of maximizing interaction between DIMP vapor and combusting metal particles and minimizing interactions with additional foreign elements, along with the spatial limitations posed by the external ports of the experimental cell, led to TDLAS being the preferred method for temperature measurements.

It is important to note that TDLAS is a path averaged measurement technique. A gradient in temperature is certainly expected along the path integrated measurement; however, the means to combat this quickly become non-trivial, and we understand a single laser TDLAS system will have limitations in spatial resolution. That being said, a higher temperature zone alone the path integrated measurement line is still incorporated in the measured spectra. Therefore, if there are significantly elevated or reduced temperatures within the measurement region, they will be captured in the final spectrum. Thus, the combination of PRiMIRS and TDLAS created an appropriate system to measure metal powder efficacy in CWA simulant defeat.

## C. DIMP decomposition

DIMP decomposition: To investigate whether evidence suggesting decomposition was observed, the following wavelength regions were identified. Region 1 consists of wavelengths between 7.4 and 7.9 $\mu$m. This region demonstrates the highest signal contribution from DIMP, whereas possible decomposition products, such as IMP and IPA, have the lowest signal contribution. Region $1^*$ encompasses wavelengths between 7.6 and 7.9 $\mu$m. This region was narrowed compared to region 1 to account for a minor absorption peak that is present in decomposition products IMP and methyl phosphonic acid (MPA) between 7.4 and 7.6 $\mu$m. While we cannot make our own reference measurement of MPA, higher resolution FTIR measurements have been performed on both compounds.[26] Region 2 spans between 7.9 and 8.3 $\mu$m and is expected to have the greatest signal contributions from decomposition products, particularly IMP but also IPA. Region 3 covers wavelengths between 9.0 and 9.3 $\mu$m. The signal observed here is also expected to have contributions from decomposition products but to a lesser degree than in region 2 particularly from IMP.

To differentiate between the impact of the background gas temperature on DIMP decomposition and the impact of the combusting powder, both homogeneous and heterogeneous experiments were conducted. As shown in Fig. 8(a), DIMP decomposition was not observed in homogeneous thermal environments up to 230 °C over several seconds. However, when (Al–8Mg):Zr powders combust in the cell and peak background temperatures rise 100 °C but stayed below 230 °C [Fig. 6(c)], the observed spectral transformations in Fig. 8 provide evidence of DIMP decomposition and product formation. The spectral transformation of interest is a measured absorbance decrease in regions corresponding to DIMP (regions 1 and $1^*$) and a measured absorbance increase in regions corresponding to decomposition products IPA and IMP (regions 2 and 3).

This is further supported by Fig. 9, which quantifies the integrated absorbance values over time for each numbered region during experimentation. Each point calculates the integrated absorbance value between the specified wavelengths on individual spectra acquired at 100 Hz and sequentially plotted. In the plots in Fig. 9, there is a brief downward spike in the calculated integrated absorbances. As noted earlier, this is due to metal powder combustion releasing infrared radiation at both mid and long wave-IR regions.[67] Upon ignition of (Al–8Mg):Zr, a temporary flash of IR light results in the live signal $I$, exceeding the background measurement $(I_o)$ taken prior to ignition. Because experimental absorption values are calculated using Eq. (2), the calculated absorbance is obscured for a brief period, resulting in an artificial drop where the data cannot be analyzed properly. This is responsible for the sharp drop in the integrated area, and these regions are not considered during analysis. The duration of this interruption, though, is less than 500 $\mu$s, and we can readily analyze the signals prior to and shortly after ignition of the powders. The decrease in the DIMP signal and the rise in the product signals following combustion are relatively constant out to 5 s and do not return to the baseline signals in regions 1 and 1* (DIMP) and regions 2 and 3 (IPA and IMP).

The Beer–Lamber law in Eq. (2) indicates that absorption is proportional to concentration for a given species. From our calibration measurements of DIMP, IPA, and IMP, we know that the strongest contributor to region 1 is DIMP and regions 2 and 3 have greater signal contributions from by-products IMP and IPA. Looking at prior studies,[25,26] we can see that an additional by-product methyl phosphonic acid (MPA) has absorption near the left edge of region 1 below 7.6 $\mu$m. Looking closer at the temporally resolved spectra, we do see slight signal increases in this region at time points well after the combustion event. This can be seen in Fig. 8, where the smaller primary DIMP peak at 7.5 $\mu$m increases at later time points (dark red) compared to earlier time points (shaded teal and orange). This same trend is not observed in the larger primary DIMP peak at 7.8 $\mu$m; the later time points (dark red) are not elevated above the prior time points (teal and orange). Since MPA could not be easily measured, we were unable to generate in-house reference spectra. Thus, to avoid concerns with MPA signal enhancement, both regions 1 (7.4–7.9 $\mu$m) and 1* (7.6–7.9 $\mu$m) were analyzed.

Prior to ignition, we note a minor but steady decrease in signal intensity over time. In Fig. 9, this phenomenon is observed between 1.0 and 1.5 s when the nichrome wire is resistively heating but has not yet ignited the metal powder. The heating filament behaves as a secondary IR source, similar to what was explained earlier with the ignited metal powders, but to a lesser degree. As the reference is taken prior to the experiment, this feature will remain in the final spectra. By monitoring the signal prior to enabling current flow to the wire, we can support this theory as this phenomenon is seen in all regions not just those pertaining to DIMP or the by-products. Immediately upon ignition, the nichrome wire breaks and is no longer a contributor to the experiment. Experiments were performed by heating the filament without the presence of igniting metal powder to confirm that DIMP decomposition is not observed from the initiation source.

In region 3, we do observe signal enhancement in both the temporally resolved spectra in Fig. 8, as well as the integrated absorbances in Fig. 9. However, noise in this region is more prevalent in both the homogeneous thermal experiment and heterogeneous particle experiment. This is a result of residing close to the bounds of our spectral region and the start of a very prominent DIMP peak. The best locations to quantify DIMP decomposition are regions 1* and 2 as the integrated areas show a notable decrease to region 1* (DIMP) and a distinct increase to region 2 (IMP + IPA).

The absolute concentrations of DIMP and potential by-products cannot be calculated from infrared spectra without measuring reference spectra containing unique signals of every decomposition product—a very difficult and convoluted process when it comes to DIMP. An ideal method for monitoring species concentration from infrared spectra is to perform a deconvolution using known constituent spectra as can be seen in the work of Senyurt *et al.*[71] This is challenging to perform with PRiMIRS due to limitations in the spectral range. With overlapping features between DIMP and by-products, as well as difficulty obtaining additional by-product reference spectra, including unstable products that cannot be obtained via a vendor, it was determined that quantitative concentration measurements of by-product species were beyond the scope of this work. However, the ability to quantitatively analyze regions 1* and 2 allows for the qualitative differentiation of varying defeat capabilities should they occur when different powder formulations are tested. Comparing Figs. 7(b) and 8(b) provides confidence that this spectral transformation is indicative of decomposition.

Laying a foundation for this capability and establishing an approach that can be developed further was the objective of this study. The development of these spectrometers and the combined system introduces a novel method for evaluating combusting metal particle efficacy in neutralizing chemical warfare agent simulants. Not only can we verify whether or not CWA simulant decomposition occurs but we can also track the background temperature rise for a particular metal powder formulation during the chemical agent defeat experiment. This offers a means to understand what underlying factors generated by metal particle combustion [nano-oxide (soot) generation, background temperature rise, particle chemistry, etc.] contribute to CWA defeat. By varying nano-oxide production for a given rise in background temperature, we seek to understand how best to optimize composite metal powder formulations to most effectively neutralize CWA simulants.

## V. CONCLUSION

This work aimed to investigate two novel diagnostic systems, PRiMIRS and TDLAS, and demonstrate whether they can assess the ability to combust composite metal powders (Al–8Mg):Zr to neutralize DIMP at sub-second timescales. PRiMIRS was operated at parameters providing a wavelength scan range of ~2 $\mu$m with a spectral resolution of 15 cm$^{-1}$, and quantitative analysis was performed between 7.4 and 9.3 $\mu$m. The scan speed was 1.3 kHz from which spectral averaging was used to produce 100 Hz spectra with SNR enhancement. TDLAS was operated at 100 Hz to match the speed at which PRiMIRS was operated for SNR enhancement, and thermal measurements were recorded between room temperature and 230 °C. Temperature resolution is calculated within ±5 °C.

Notably, in homogeneous thermal environments up to 230 °C, DIMP decomposition was not observed over several seconds, whereas in a heterogeneous combusting metal powder environment, spectral transformations indicative of DIMP defeat were observed within two seconds. We provide a comprehensive evaluation of these diagnostic systems by utilizing a series of calibration measurements performed with multiple compounds, simulations, temperature set-points, and alternative measurement techniques. These novel systems are highly tunable to other experimental demands and may be leveraged broadly toward other materials science investigations. Future analyses of DIMP decomposition will combine these two spectrometers into a unified system for simultaneous measurements of DIMP and vapor temperature. This will allow for a more comprehensive analysis as we aim to understand how varying metal powder chemistry affects DIMP decomposition.

## ACKNOWLEDGMENTS

This work was supported by the Department of Defense, Defense Threat Reduction Agency (DTRA) under the MSEE URA, Grant No. HDTRA1-20-2-0001.

## AUTHOR DECLARATIONS

### Conflict of Interest

The authors have no conflicts to disclose.

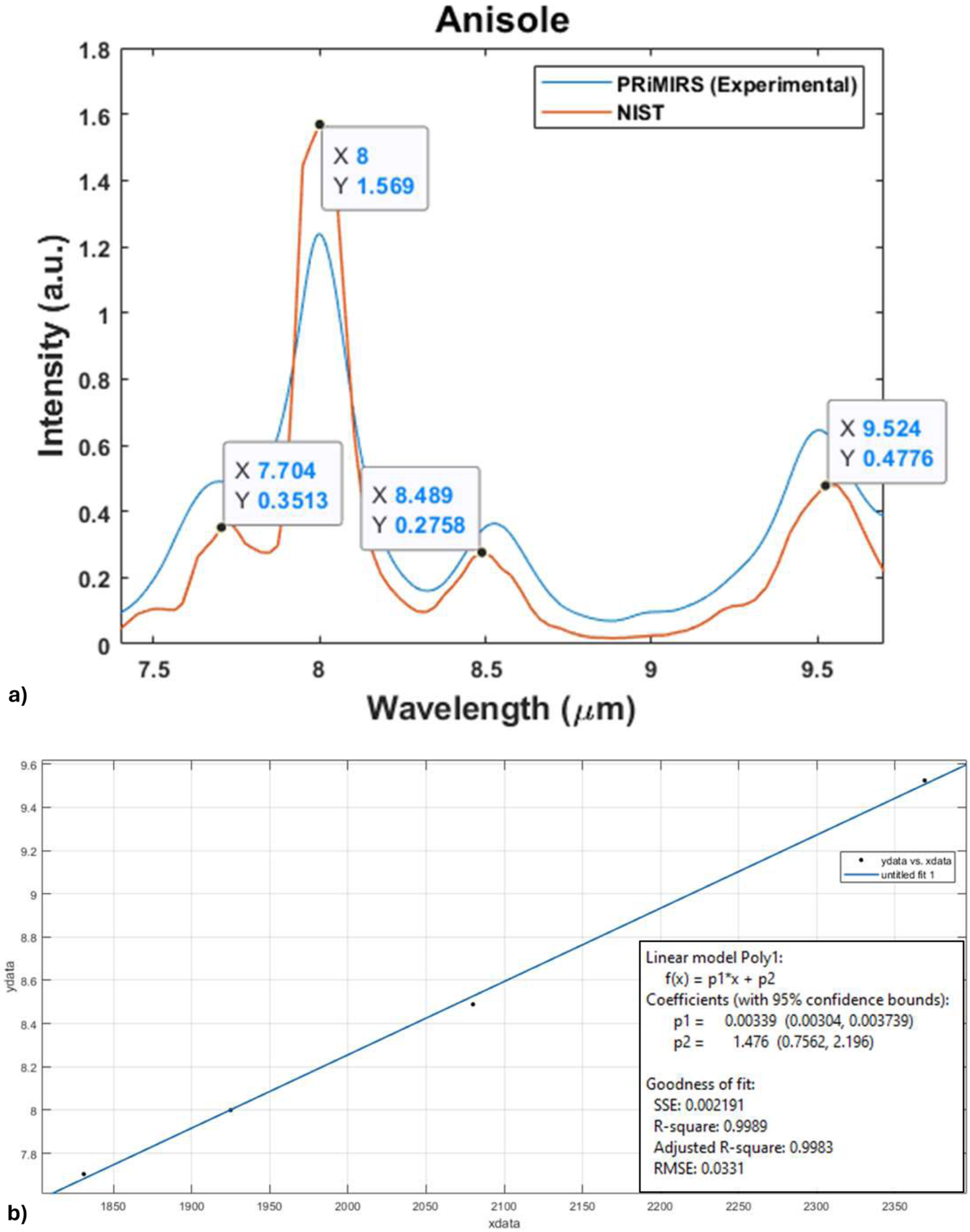

**FIG. 10.** Reference wavelengths are identified using spectra available on the NIST database. Peak maxima from experimental data are plotted against reference wavelengths, and linear fitting is performed in the Matlab curve fitting toolbox. The following peaks for each compound are used: anisole—7.70, 8.00, 8.49, and 9.52 $\mu$m, ethyl benzoate—7.85, 8.49, and 9.01 $\mu$m, and $NH_3$—8.77, 8.93, 9.07, 9.24, 9.36, 9.54, and 9.68 $\mu$m.

## Author Contributions

**Preetom Borah**: Conceptualization (lead); Data curation (lead); Formal analysis (lead); Investigation (lead); Methodology (lead); Software (lead); Validation (lead); Visualization (lead); Writing – original draft (lead); Writing – review & editing (supporting). **Milad Alemohammad**: Methodology (supporting). **Mark Foster**: Funding acquisition (supporting); Methodology (supporting); Supervision (supporting); Writing – review & editing (supporting). **Timothy P. Weihs**: Funding acquisition (lead); Project administration (lead); Supervision (lead); Writing – review & editing (lead).

## DATA AVAILABILITY

The data that support the findings of this study are available within the article. Additional data are available from the corresponding authors upon reasonable request.

## APPENDIX: SUPPLEMENTARY FIGURES FOR CALIBRATION AND ERROR ASSESSMENT

### 1. Linear fit wavelength calibration

Reference wavelengths are identified using spectra available on the NIST database (see Fig. 10).

### 2. Additional minimizations between experimental and simulated data

Figure 11 shows additional minimizations are shown between experimental and simulated spectra for the thermal profile of (Al–8Mg):Zr ignited at room temperature shown in Fig. 6(a).

### 3. Associated error bars for dynamic events

Table I shows associated error bars for dynamic events. Error bars for dynamic events are generated from taking the standard deviation from a region prior to initiating the combustion event.

**TABLE I.** Error bars for dynamic events are generated from taking the standard deviation from a region prior to initiating the combustion event. Individual combustion events are unique and will have inherent variance from run to run. Therefore, to best account for associated error, calculating the variation or noise during a stable thermal period is appropriate. Error bars are connected and shown as shaded regions in each respective plot for visual clarity due to the density of data points measured from experimental acquisition speed.

| Table of associated error bars calculated from the standard deviation of stable regions | |
|---|---|
| Figure Nos. | Associated error from STD |
| Figure 6(a) | 2.67 n = 100 |
| Figure 6(b) 18C TDLAS | 2.82 n = 100 |
| Figure 6(b) 115C TDLAS | 2.32 n = 100 |
| Figure 6(b) 150C TDLAS | 5.16 n = 100 |
| Figure 6(c) bottom | 4.67 n = 100 |
| Figure 6(c) middle | 3.11 n = 100 |
| Figure 6(c) top | 10.01 n = 100 |
| Figure 9(a) | 0.0027 n = 75 |
| Figure 9(b) | 0.0018 n = 75 |
| Figure 9(c) | 0.0012 n = 75 |
| Figure 9(d) | 0.0025 n = 75 |

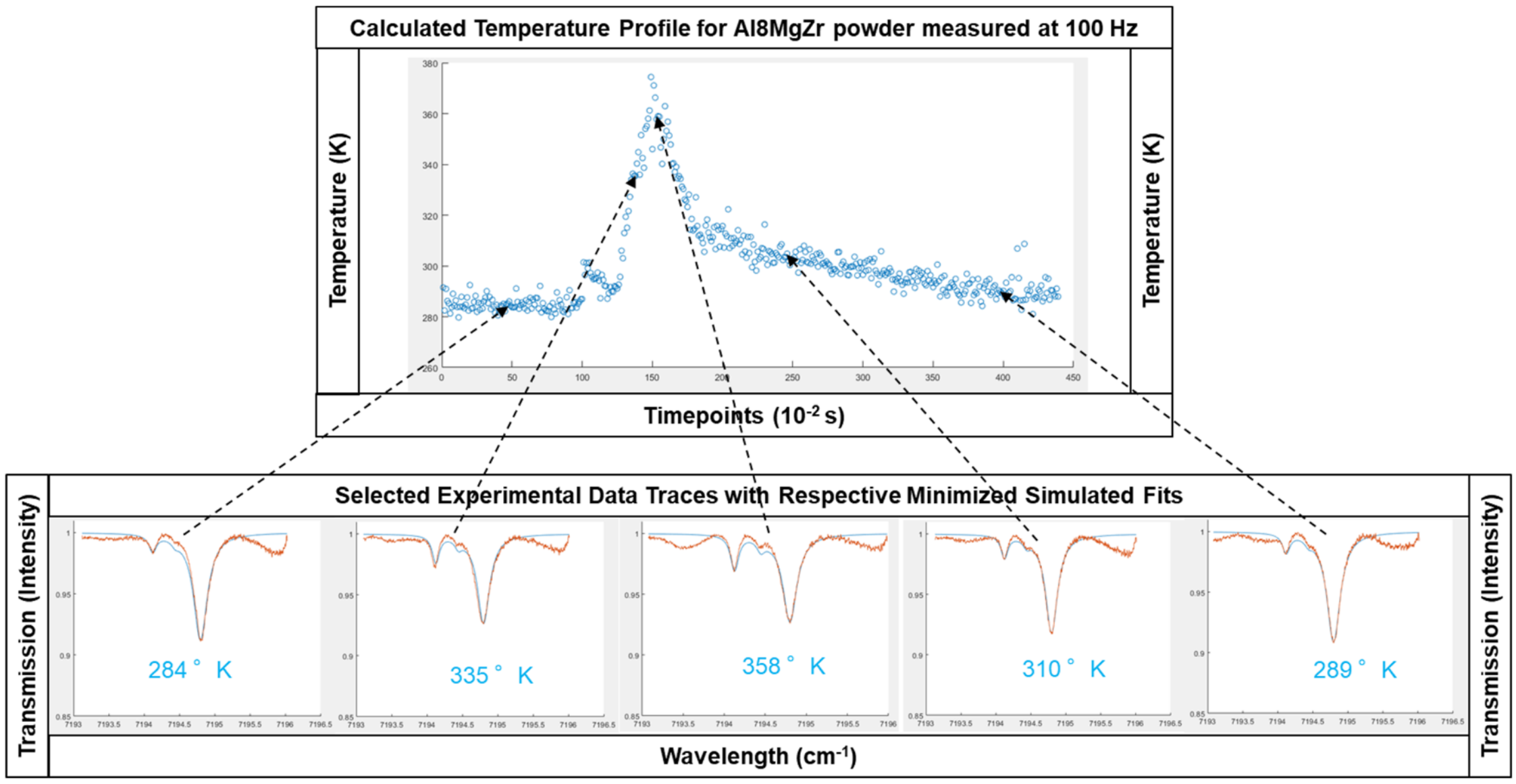

**FIG. 11.** Additional minimizations are shown between experimental and simulated spectra for the thermal profile of (Al–8Mg):Zr ignited at room temperature shown in Fig. 6(a).